\documentclass[11pt, a4paper]{article}
\usepackage{amssymb}
\usepackage{amsmath}

\numberwithin{equation}{section}

\usepackage{graphicx,color}
\usepackage{wrapfig,framed}

\newtheorem{theorem}{Theorem}

\newtheorem{lemma}[theorem]{Lemma}

\newtheorem{cor}[theorem]{Corollary}

\newtheorem{remark}[theorem]{Remark}

\def \= {\;=\;}
\def \+ {\,+\,}

\newcommand{\eqa}{\begin{eqnarray}}
\newcommand{\eeqa}{\end{eqnarray}}
\newcommand{\beq}{\begin{equation}}
\newcommand{\eeq}{\end{equation}}
\newcommand{\nn}{\nonumber}

\newcommand{\pal}{\partial}

\newcommand{\CC}{\mathbb{C}}

\newcommand{\CW}{{\mathcal W}}

\newcommand{\tr}{{\rm tr}}

\newcommand{\bt}{{\bf t}}

\newcommand{\ba}{{\bf a}}
\newcommand{\bb}{{\bf b}}
\newcommand{\bc}{{\bf c}}

\newcommand{\bv}{{\bf V}}
\newcommand{\bu}{{\bf u}}

\newcommand{\epf}{$\quad$\hfill
\raisebox{0.11truecm}{\fbox{}}\par\vskip0.4truecm}

\usepackage{array}
\newcolumntype{M}[1]{>{\centering\arraybackslash}m{#1}}
\newcolumntype{R}[1]{>{\raggedleft\arraybackslash}m{#1}}
\newcolumntype{N}{@{}m{0pt}@{}}

\def\XXint#1#2#3{{\setbox0=\hbox{$#1{#2#3}{\int}$}
\vcenter{\hbox{$#2#3$}}\kern-.5\wd0}}

\title{Approximating tau-functions by theta-functions}\par
\author{{Boris DUBROVIN}
}
\date{}
\begin{document}
\maketitle

\begin{abstract} We prove that the logarithm of an arbitrary tau-function of the KdV hierarchy can be approximated, in the topology of graded formal series by the logarithmic expansions of hyperelliptic theta-functions of finite genus, up to at most quadratic terms. As an example we consider theta-functional approximations of the Witten--Kontsevich tau-function.
\end{abstract}

\setcounter{equation}{0}
\setcounter{theorem}{0}
\section{Introduction}\par


Consider the algebra $\CW=\mathbb C[\ba,\bb,\bc]$ of polynomials in infinite number of variables
$$
\ba=(a_1, a_2, \dots), \quad \bb=(b_1, b_2, \dots), \quad \bc=(c_1, c_2, \dots).
$$
Define a gradation on $\CW$ by
\beq\label{grad}
\deg a_i=2i+1, \quad \deg b_i=\deg c_i=2i.
\eeq
Introduce the matrix-valued formal Laurent series in $1/z$
\beq\label{wser}
W(z)\equiv W(\ba,\bb,\bc;z)=\left(\begin{array}{cc} 0& 1\\z+c_1 & 0\end{array}\right)+\sum_{i\geq 1}\left(\begin{array}{cc} a_i & b_i\\ c_{i+1} & -a_i\end{array}\right)\frac1{z^i}
\eeq
and define a Poisson algebra structure on $\CW$ by means of  the following Poisson bracket
\beq\label{braser}
\left\{ W(z_1)\underset{,}{\otimes} W(z_2)\right\} =\left(\left[ R(z_1-z_2), W(z_1)\otimes {\bf 1}+{\bf 1} \otimes W(z_2)\right]\right)_-+\left[\Delta R , W(z_1)\otimes {\bf 1}-{\bf 1} \otimes W(z_2)\right]
\eeq
where
\beq\label{braser1}
R(z)=\frac{P}{z}
\eeq
is the standard $r$-matrix,
$$
P:\mathbb C^2\otimes \mathbb C^2 \to \mathbb C^2\otimes \mathbb C^2, \quad P(x\otimes y)=y\otimes x
$$
and
\beq\label{braser2}
\Delta R=E_{21}\otimes E_{21}\quad \text{with}\quad E_{21}=\left(\begin{array}{cc} 0 & 0\\ 1 & 0\end{array}\right).
\eeq
For any pair of homogeneous elements $f,\, g\in\CW$ one has
\beq\label{bradeg}
\deg \{ f, g\}=\deg f+\deg g-3.
\eeq
The annihilator of the Poisson bracket coincides with the subring $\mathbb C[b_1+c_1]$.

We will now introduce an infinite family of derivations on the algebra $\CW$. Let
\beq\label{qser}
Q(z)=-\det W(z)=z+\sum_{i\geq 1}\frac{q_i}{z^{i-1}}.
\eeq
Define an infinite sequence of polynomials $H_n\in\CW$, \quad $n\geq -1$ using coefficients of the following series
\beq\label{hamser}
1+\frac12\sum_{n\geq -1}\frac{H_n}{z^{n+2}}=\sqrt{\frac{Q(z)}{z}}.
\eeq
The polynomial $H_{-1}=b_1+c_1$ is the Casimir of the Poisson bracket.
The derivations $\pal_n$ are defined as Hamiltonian vector fields
\beq\label{derser}
\pal_nf=\{  H_n,f\}\quad\forall\, f\in\CW, \quad  n\geq 0.
\eeq
Explicitely
\eqa
&&
\pal_0=\left( c_2-b_2+b_1^2-b_1 c_1\right)\frac{\pal}{\pal a_1}-2a_1\frac{\pal}{\pal b_1}+2 a_1 \frac{\pal}{\pal c_1}+
\nn\\
&&
+\left( c_3-b_3 +b_2(b_1-c_1)\right)\frac{\pal}{\pal a_2}-2 a_2\frac{\pal}{\pal b_2}+2\left( a_2 +a_1(c_1-b_1)\right)\frac{\pal}{\pal c_2}+\dots
\nn
\eeqa
\eqa
&&
\pal_1=\left( c_3-b_3+\frac12 b_2(3b_1-c_1)-\frac12 c_2(b_1+c_1)-\frac12 b_1(b_1^2-c_1^2)\right)\frac{\pal}{\pal a_1}+\left(-2 a_2+a_1(b_1+c_1)\right)\frac{\pal}{\pal b_1}+
\nn\\
&&
+\left(2 a_2 +a_1(b_1+c_1)\right)\frac{\pal}{\pal c_1}+\left( c_4-b_4+\frac12(b_3+c_3)(b_1-c_1)-\frac12 b_2(2c_2-2b_2+b_1^2-c_1^2
\right)\frac{\pal}{\pal a_2}+
\nn\\
&&
+\left(-2a_3+a_2(c_1-b_1)+2a_1b_2\right)\frac{\pal}{\pal b_2} +\left( 2a_3 +a_2(c_1-b_1)-2a_1b_2+a_1(b_1^2-a_1^2)\right)\frac{\pal}{\pal c_2}+\dots
\nn
\eeqa
etc. Note that the Hamiltonian $H_n$ is a graded homogeneous polynomial of degree $2n+4$. Thus the derivation $\pal_n$ increases the degree by $2n+1$. 

We will now derive a ``commutator representation" for the action of the vector fields $\pal_n$ on $W(z)=W(\ba,\bb,\bc,z)$. To this end introduce another matrix-valued series
\eqa\label{mser}&&
M(z)\equiv M(\ba,\bb,\bc;z)=\frac{W(z)}{\sqrt{Q(z)/z}}=\left(\begin{array}{cc}0 & 1\\ z& 0\end{array}\right)+\sum_{i\geq 0}\left(\begin{array}{cc}\tilde a_i & \tilde b_i\\ \tilde c_{i+1} & -\tilde a_i\end{array}\right)\frac1{z^i}
\\
&&
\tilde a_i, ~\tilde b_i, ~ \tilde c_i\in\CW, \quad \tilde a_0=\tilde b_0=0, \quad \tilde c_1=\frac{c_1-b_1}2. 
\nn
\eeqa
Note that
$$
\det M(z)=-z.
$$

\begin{lemma} \label{lem11} For any $n\geq 0$ the following equation holds true
\beq\label{laxser}
\pal_nW(z)=\left[ U_n(z), W(z)\right], \quad n=0,\, 1, \, 2, \dots
\eeq
where
\beq\label{user}
U_n(z)=\left[ z^{n+1} M(z)\right]_+-\left(\begin{array}{cc} 0 & 0\\ \tilde b_{n+1} & 0\end{array}\right).
\eeq 
\end{lemma}

\begin{cor} \label{cor12} The derivations $\pal_n$ commute pairwise.
\end{cor}



\begin{remark} \label{rem424} One can consider in a similar way the graded algebra $\CW_g$ of polynomials in $3g+1$ variables
$$
\CW_g=\mathbb C[a_1,\dots, a_g, b_1, \dots, b_g, c_1, \dots, c_{g+1}].
$$
A Poisson bracket on $\CW_g$ defined by a formula similar to \eqref{braser} first appeared in \cite{NS}. However it \emph{differs} from the one induced by restriction of \eqref{braser} wrt the natural embedding
$$
\CW_g\subset \CW, \quad W(z) \mapsto \frac1{z^g} W(z).
$$
\end{remark}

Due to the commutativity of derivations $[\pal_n,\pal_m]=0$ for an arbitrary triple of sequences of complex numbers $a_i^0$, $b_i^0$, $c_i^0$, $i\geq 1$ there exists a unique common solution
$$
a_i(\bt), ~b_i(\bt), ~c_i(\bt)\in\mathbb C[[t_0, t_1, \dots]], \quad i=1, \, 2, \dots
$$
to the following infinite system of Hamiltonian differential equations
\beq\label{kdvser}
\frac{da_i}{dt_k}=\pal_k a_i, \quad \frac{db_i}{dt_k}=\pal_k b_i, \quad \frac{dc_i}{dt_k}=\pal_k c_i, \quad i=1, \, 2, \dots, \quad k=0, \, 1, \dots
\eeq
satisfying the initial conditions
$$
a_i(0)=a_i^0, ~b_i(0)=b_i^0, ~c_i(0)=c_i^0, \quad i=1, \, 2, \dots
$$
We will now establish a relationship between this infinite system of commuting polynomial ODEs with the Korteweg--de Vries (KdV) hierarchy of PDEs
\eqa
&&
u_{t_0}=u_x
\nn\\
&&\label{kdv}
u_{t_1}=3 u\, u_x+\frac14 u_{xxx}
\\
&&
u_{t_2}=\frac{15}2u^2 u_x +\frac52\left(2 u_x u_{xx} +u\, u_{xxx}\right)+\frac1{16} u^{(5)}
\nn
\eeqa
etc. that can be represented in the Lax form
$$
\frac{\pal}{\pal t_n}L=\left[\left(L^{\frac{2n+1}2}\right)_+, \, L\right], \quad L=\pal_x^2+2u.
$$

Consider the matrix-valued series
$$
W^0(z)=\left(\begin{array}{cc} 0& 1\\z+c_1^0 & 0\end{array}\right)+\sum_{i\geq 1}\left(\begin{array}{cc} a_i^0 & b_i^0 \\ c_{i+1}^0  & -a_i^0 \end{array}\right)\frac1{z^i}.
$$
For any $N\geq 2$ define a collection of numbers $F^0_{k_1\dots k_N}$ labelled by indices $k_1$, \dots, $k_N=0,\, 1, \dots$ by the following generating series
\eqa\label{tauser}
&&
\sum_{k_1, \dots, k_N} \frac{F^0_{k_1\dots k_N}} {z_1^{k_1+1}\dots z_N^{k_N+1}}=
\\
&&
=-\frac1{N}\frac1{\sqrt{\frac{Q(z_1)}{z_1}}\dots\sqrt{\frac{Q(z_N)}{z_N}}}\sum_{s\in S_N} \frac{\tr \left[W^0(z_{s_1})\dots W^0(z_{s_N})\right]}{(z_{s_1}-z_{s_2})\dots (z_{s_{N-1}}-z_{s_N})(z_{s_N}-z_{s_1})}-\delta_{N,2}\frac{z_1+z_2}{(z_1-z_2)^2}.
\nn
\eeqa
Define a function $F(\bt)\in \mathbb C[[\bt]]$ by its formal Taylor series expansion
\beq\label{tauser1}
F(\bt) =\sum_{N\geq 2}\frac1{N!}\sum_{k_1, \dots, k_N} F^0_{k_1\dots k_N} t_{k_1}\dots t_{k_N}.
\eeq

\begin{theorem} \label{thm15} For arbitrary initial conditions $a_i^0$, $b_i^0$, $c_i^0$ the function
$$
u(\bt)=\frac{\pal^2 F(\bt)}{\pal t_0^2}
$$
satisfies equations of the KdV hierarchy. The tau-function of this solution  is equal to
$$
\tau(\bt)=e^{F(\bt)}
$$
up to a transformation
$
\tau(\bt)\mapsto e^{\alpha+\sum \beta_i t_i}\tau(\bt)
$
with some constant coefficients $\alpha$, $\beta_i$.
Any solution $u(\bt)\in \mathbb C[[\bt]]$ of the KdV hierarchy along with its tau-function can be obtained by this procedure.
\end{theorem}

\begin{remark} Changing $W(z)$ by a scalar factor
$$
W(z)\mapsto f(z)\, W(z),\quad f(z)=1+\sum_{i\geq1}\frac{f_i}{z^i}\in\CC[[1/z]]
$$
does not change the KdV solution.
\end{remark}

Note that the polynomial $F_{i_1\dots i_N}(\ba,\bb,\bc)=\pal^N\log\tau/\pal t_{i_1}\dots\pal t_{i_N}$ defined by eq. \eqref{tauser} is graded homogeneous of the degree $2(i_1+\dots+i_N)+N$. Therefore it belongs to the subring $\CW_g=\mathbb C[a_1, \dots, a_g, $ $b_1,\dots,b_g,c_1,\dots, c_{g+1}]$ for
$$
g=i_1+\dots+i_N+\left[\frac{N}2\right].
$$
We can reformulate this observation in the following way. 

Let
$$
f(\bt)=\sum_{n\geq 0}\frac1{n!}\sum_{i_1,\dots, i_n}f_{i_1\dots i_n}t_{i_1}\dots t_{i_n}\in \mathbb C[[\bt]]
$$
be a formal series of an infinite number of variables $\bt=(t_0, t_1, \dots)$. For a given number $m>0$ denote
\beq\label{trunc}
\left[f(\bt)\right]_m=\sum_{n\geq 0}\frac1{n!}\sum_{2(i_1+\dots+ i_n)+n\leq m}f_{i_1\dots i_n}t_{i_1}\dots t_{i_n}
\eeq
the $m$\emph{-truncation} of the series. 

\begin{cor} Let $\tau(\bt)$ be the tau-function of an arbitrary solution $u(\bt)\in\mathbb C[[\bt]]$ of the KdV hierarchy. Then for any $m>0$ there exists a hyperelliptic curve $C$ of genus less or equal than $\left[\frac m2\right]$ and a point $\bu_0\in J(C)\setminus\left(\Theta\right)$ in the Jacobian such that
$$
\left[\log\tau(\bt)\right]_m
=\left[ \log\theta \left(\sum_{2i+1\leq m}t_i \bv^{(i)}-\bu_0\right)\right]_m+\alpha +\sum_{2i+1\leq m} \beta_i t_i +\frac12\sum_{2(i+j+1)\leq m}\gamma_{ij} t_i t_j
$$
for some constants $\alpha$, $\beta_i$, $\gamma_{ij}$.
\end{cor}

\begin{remark} It can happen that the curve $C$ is singular. In that case one has to deal with the generalized Jacobian and the corresponding analogue of theta-function.
\end{remark}

Our last remark is about a block-triangular invertible change of variables between the generators of the algebra $\CW$ and the jet variables $u$, $u_x$, \dots depending on the constants of motion $q_i$ (see eq. \eqref{qser})
\beq\label{blockchange}
\{ a_1, b_1, c_1, \dots, a_n, b_n, c_n\} \leftrightarrow \{ f_2, \dots, f_{2n+1}, q_1, \dots , q_n\}
\eeq
for every $n\geq 1$. Here
$$
f_2=\frac12(b_1-c_1)=\left(\log\tau\right)_{00}=u,\quad f_3 =\left(\log\tau\right)_{000}=u_x,\dots, f_{2n+1}=\pal_0^{2n+1}\log\tau=u^{(2n-1)}.
$$
Explicitly
$$
a_1=-\frac12 u_x,\quad b_1=u+\frac12 q_1,\quad c_1=-u+\frac12 q_1
$$
$$
a_2=-\frac18 u_{xxx}-\frac32 u\, u_x -\frac14 q_1 u_x, \quad b_2=\frac14 u_{xx}+\frac32 u^2 +\frac12 q_1 u +\frac{q_2}2 -\frac{q_1^2}2,\quad c_2=-\frac14 u_{xx}-\frac12 u^2 -\frac12 q_1 u +\frac{q_2}2 -\frac{q_1^2}2
$$
etc.
After such a change the matrix polynomials $U_n(z)$ take the familiar form
$$
U_0=\left(\begin{array}{cc} 0 & 1\\ z-2u & 0\end{array}\right), \quad U_1=\left( \begin{array}{cc} -\frac{u_x}2 & z+u\\
z^2 -u\, z -\frac{u_{xx}}2-2u^2 & \frac{u_x}2\end{array}\right)
$$
etc. Observe that the coefficients of these matrix polynomials depend only on the jet variables but not on the integration constants $q_i$.

\medskip

The constructions of the present paper can be generalized to other spaces of matrix-valued series. Moreover they can be extended to series with coefficients in an arbitrary simple Lie algebra. This will be done in a separate publication.

\setcounter{equation}{0}
\setcounter{theorem}{0}
\section{Proofs}\par

We begin with the proof of Lemma \ref{lem11}. Let us rewrite the Poisson bracket \eqref{braser} in coordinates,
\eqa\label{braser3}
&&
\{a(z),a(w)\}=0, \quad \{a(z),b(w)\}=-\frac{b(z)-b(w)}{z-w}, \quad \{a(z),c(w)\}=\frac{c(z)-c(w)}{z-w}-b(z)
\nn\\
&&
\{b(z),b(w)\}=0, \quad \{b(z),c(w)\}=-2\frac{a(z)-a(w)}{z-w}, \quad \{c(z),c(w)\}=2[a(z)-a(w)]
\eeqa
where we denote
$$
a(z)=\sum_{i\geq 1}\frac{a_i}{z^i}, \quad b(z)=\sum_{i\geq 1}\frac{b_i}{z^i}, \quad c(z)=\sum_{i\geq 1}\frac{c_i}{z^{i-1}}.
$$
Using \eqref{braser3} obtain
$$
\left\{  Q(w),W(z)\right\}\equiv \left(\begin{array}{lr} \{ Q(w),a(z)\} & ~\{Q(w),b(z)\}\\ \{Q(w),c(z),\} & -\{ Q(w),a(z)\}\end{array}\right)=\frac1{w-z}\left[W(w),W(z)\right]-b(w)\left[E_{21},W(z)\right].
$$
Thus
\beq\label{hw}
\left\{\sum_{n\geq -1}\frac{H_n}{w^{n+1}},W(z)\right\}=2w\left\{\sqrt{\frac{Q(w)}{w}},W(z)\right\}=\frac{1}{w-z}\left[M(w),W(z)\right]-\tilde b(w)\left[E_{21},W(z)\right]
\eeq
where
$$
\tilde b(w)=\sum_{i\geq 1}\frac{b_i}{z^i}.
$$
Collecting the coefficient of $w^{-n-1}$ in eq. \eqref{hw} we obtain \eqref{laxser}. \epf

{\it Proof} of Corollary \ref{cor12}. Due to the commutator structure of the rhs of eq. \eqref{hw} we deduce that
$$
 \left\{\sum_{n\geq -1}\frac{H_n}{w^{n+1}},\det W(z)\right\}=0
$$
hence
$$
\left\{\sum_{n\geq -1}\frac{H_n}{w^{n+1}},\sum_{m\geq -1}\frac{H_m}{z^{m+1}}\right\}=0.
$$\epf

{\it Proof} of Theorem \ref{thm15}. Introduce a $\bt$-dependent matrix series
$$
W(\bt,z):=W\left(\ba(\bt), \bb(\bt), \bc(\bt),z\right)
$$
and
$$
M(\bt,z):=M\left(\ba(\bt), \bb(\bt), \bc(\bt),z\right)=\frac{W(\bt,z)}{\sqrt{Q(z)/z}}.
$$
Due to Lemma \ref{lem11} the matrix series $W(\bt,z)$ satisfies
$$
\frac{d}{dt_n}W(\bt,z)=\left[U_n(\bt,z),W(\bt,z)\right],\quad n\geq 0
$$
where the matrix polynomials $U_n(\bt,z)$ are obtained from the series $M(\bt,z)$ by the construction of eq. \eqref{user}. In particular,
\beq\label{u0}
U_0(\bt,z)=\left(\begin{array}{cc} 0 & 1\\ z-2u(\bt) & 0\end{array}\right), \quad u(\bt):=\frac12(b_1(\bt)-c_1(\bt)).
\eeq
Similar equations hold true for $M(\bt,z)$
$$
\frac{d}{dt_n}M(\bt,z)=\left[U_n(\bt,z),M(\bt,z)\right],\quad n\geq 0
$$
Because of the commutativity of the derivations the matrix polynomials $U_n(\bt,z)$ satisfy the ``zero curvature equations"
$$
\left[\frac{d}{dt_n}-U_n(\bt,z), \frac{d}{dt_m}-U_m(\bt,z)\right]=0\quad \forall~m, \, n\geq 0
$$
that imply equations of the KdV hierarchy \eqref{kdv}
for the function $u(\bt)$.

To compute the tau-function of this solution, according to the recipe of \cite{BDY1} we have to find the so-called matrix resolvent, i.e. the unique matrix series $R(\bt,z)$ of the form
$$
R(\bt,z)=\left(\begin{array}{cc}0 & 1\\z-u(\bt) & 0\end{array}\right)+{\mathcal O}\left(\frac1{z}\right)
$$
satisfying the equations
$$
\frac{d}{dt_n}R(\bt,z)=\left[ U_n(\bt,z), R(\bt,z)\right], \quad n\geq 0
$$
along with the normalization
$$
\tr \,R(\bt,z)=0, \quad\det R(\bt,z)=-z.
$$
Then the $N$-th order logarithmic derivatives
$$
F_{k_1\dots k_N}(\bt):=\frac{\pal^N\log\tau(\bt)}{\pal t_{k_1}\dots\pal t_{k_N}}
$$
of the tau-function of this solution for any $N\geq 2$ can be determined from the following generating series
\beq\label{resser}
\sum_{k_1, \dots, k_N} \frac{F_{k_1\dots k_N}(\bt)} {z_1^{k_1+1}\dots z_N^{k_N+1}}
=-\frac1{N}\sum_{s\in S_N} \frac{\tr \left[R(\bt, z_{s_1})\dots R(\bt,z_{s_N})\right]}{(z_{s_1}-z_{s_2})\dots (z_{s_{N-1}}-z_{s_N})(z_{s_N}-z_{s_1})}-\delta_{N,2}\frac{z_1+z_2}{(z_1-z_2)^2}.
\eeq
Clearly the matrix series $M(\bt,z)$ satisfies all these conditions, so
$R(\bt,z)=M(\bt,z)$. Therefore the logarithmic derivatives of the tau-function of the solution $u(\bt)$ are given by
\eqa
&&
\sum_{k_1, \dots, k_N} \frac{F_{k_1\dots k_N}(\bt)} {z_1^{k_1+1}\dots z_N^{k_N+1}}=
\nn\\
&&
=-\frac1{N}\frac1{\sqrt{\frac{Q(z_1)}{z_1}}\dots\sqrt{\frac{Q(z_N)}{z_N}}}\sum_{s\in S_N} \frac{\tr \left[W(\bt,z_{s_1})\dots W(\bt, z_{s_N})\right]}{(z_{s_1}-z_{s_2})\dots (z_{s_{N-1}}-z_{s_N})(z_{s_N}-z_{s_1})}-\delta_{N,2}\frac{z_1+z_2}{(z_1-z_2)^2},
\nn
\eeqa
$N\geq 2$. For $\bt=0$ it reduces to eq. \eqref{tauser1}. This completes the proof of the first part of the theorem.

Converesely, let $u(\bt)\in\mathbb C[[\bt]]$ be a solution to the equations of the KdV hierarchy. Let $R(\bt,z)\in Mat_2\left(\mathbb C[[\bt]]\right)\otimes z \mathbb C[[1/z]]$ be the matrix resolvent of this solution. Put $W^0(z):=R(0,z)$. Then $Q(z)=z$ so eq.  \eqref{resser} at $\bt=0$ coincides with \eqref{tauser1}. \epf

\setcounter{equation}{0}
\setcounter{theorem}{0}
\section{Examples}\par

\subsection{Logarithmic expansions of hyperelliptic theta-functions}\par

Let us briefly revisit the finite-gap case within the general framework. Consider the subspace of matrix series that truncate at the $g$-th term,
$$
W(z)\equiv W(\ba,\bb,\bc;z)=\left(\begin{array}{cc} 0& 1\\z+c_1 & 0\end{array}\right)+\sum_{i= 1}^g\left(\begin{array}{cc} a_i & b_i\\ c_{i+1} & -a_i\end{array}\right)\frac1{z^i}.
$$
It corresponds to the subalgebra
$$
\CW_g=\mathbb C[a_1,\dots, a_g, b_1, \dots, b_g, c_1, \dots, c_{g+1}]\subset \CW
$$
(we use the same notation as in Remark \ref{rem424}). Using the degree counting \eqref{bradeg} it is easy to verify that $\CW_g$ is a Poisson subalgebra wrt the Poisson bracket \eqref{braser}. The annihilator of the restriction of \eqref{braser} onto $\CW_g$ is generated by $g+1$ Casimirs $q_1=b_1+c_1$ and
$$
q_{g+k+1}=\sum_{i=k}^g( a_i a_{g+k-i}+b_ic_{g+k+1-i}), \quad k=1, \dots, g.
$$
The Darboux coordinates $x_1, \dots, x_g$, $p_1$, \dots, $p_g$, 
$$
\{x_i,x_j\}=\{p_i, p_j\}=0,\quad \{p_i,x_j\}=\delta_{ij}
$$
on the symplectic leaves
$$
q_1=q_1^0,~ q_{g+2}=q_{g+2}^0, \dots, q_{2g+1}=q_{2g+1}^0
$$
are obtained by the standard, see e.g. \cite{NV}, up to a change $z\mapsto 1/z$ procedure by taking the $z$- and $w$-coordinates of poles of the eigenvector of the matrix $W(z)$. I.e., $x_1$, \dots, $x_g$ are roots of the equation
$$
1+\frac{b_1}{z}+\dots +\frac{b_g}{z^g}=0
$$
and
$$
p_i=-\left(\frac{a_1}{x_i}+\dots+\frac{a_g}{x_i^g}\right), \quad i=1, \dots, g.
$$

The corresponding solutions to the KdV hierarchy are often called \emph{finite-gap} or \emph{algebro-geometric} solutions. For them the tau-function coincides with the hyperelliptic theta-function up to multiplication by exponential of a quadratic polynomial. Let us recall this construction. It is convenient to change
$$
W(z)\mapsto z^g W(z)
$$
in order to deal with matrix polynomials
of the form
\eqa\label{2gon}
&&
W(z)=\left(\begin{array}{cc}  a(z) & b(z) \\
c(z) & -a(z)\end{array}\right)
\\
&&
a(z)=a_1 z^{g-1}+\dots +a_g, \quad b(z)=z^g+b_1 z^{g-1}+\dots+b_g, \quad c(z)=z^{g+1}+c_1 z^g+\dots+c_{g+1}.
\nn
\eeqa
Assuming the roots of
$$
Q(z)=-\det W(z)=z^{2g+1}+q_1 z^{2g}+\dots+q_{2g+1}
$$
to be pairwise distinct we obtain a hyperelliptic curve $C$
\beq\label{hyperell}
w^2=z^{2g+1}+q_1 z^{2g}+\dots+q_{2g+1}
\eeq
of genus $g$ with one branch point $P_\infty$ at infinity. The fibers of the natural fibration 
$$
\begin{array}{c}\{ {\rm space~of~matrix~polynomials}~W(z)=\left(\begin{array}{cc}0 & z^g\\ 
z^{g+1}+c_1 z^g & 0\end{array}\right)+\dots+ \left(\begin{array}{cc} a_g & b_g\\ c_{g+1} & -a_i\end{array}\right)\} \\
\Big\downarrow
\\ 
\{ {\rm space~of~hyperelliptic~ curves}~w^2=z^{2g+1}+\dots+q_{2g+1}\}
\end{array}
$$
\beq\label{fibra}
W(z) \mapsto \det(W(z)-w\cdot{\bf 1})=w^2-Q(z)
\eeq
are isomorphic to the affine Jacobians $J(C)\setminus (\Theta)$ of the curves. Here $(\Theta)\subset J(C)$ is the theta divisor.  For $g=1$ it comes from an easy calculation. For $g\geq 2$ it was first observed in \cite{DN74}, see also the book \cite{Mum}. The correspondence between the matrices in the fibers of the fibration \eqref{fibra} can be established by the map
$$
W(z) \mapsto \text{line bundle over}~C~\text{of eigenvectors of}~W(z).
$$
The degree of the line bundle equals $g+1$. It is convenient to choose a representative in the class of linear equivalence in the form $D_0+P_\infty$ where
\beq\label{div1}
D_0=Q_1+\dots +Q_g, \quad Q_j=(z_j,w_j)\in C\setminus P_\infty
\eeq
is a nonspecial positive divisor of degree $g$ defined by the equations
\beq\label{div2}
b(z_j)=0, \quad w_j=-a(z_j), \quad j=1, \dots, g.
\eeq
Applying to the divisor $D_0$ the Abel--Jacobi map one obtains a point $\bu_0\in J(C)\setminus (\Theta)$. Explicitely,
\beq\label{div3}
\bu_0=\sum_{j=1}^g\left(\int_{P_\infty}^{Q_j}\omega_1,\dots, \int_{P_\infty}^{Q_j}\omega_g\right)-{\boldsymbol \varpi}
\eeq
where
\beq\label{hyperholo}
\omega_i=(\alpha_{i1}z^{g-1}+\dots+\alpha_{ig})\frac{dz}{2w}, \quad i=1, \dots, g
\eeq
are the normalized holomorphic differentials,
$$
\oint_{a_j}\omega_i=2\pi\sqrt{-1}\,\delta_{ij}
$$
for some basis $a_1$, \dots, $a_g$, $b_1$, \dots, $b_g\in H_2(C,\mathbb Z)$,
$$
a_i\circ a_j=b_i\circ b_j=0, \quad a_i\circ b_j=\delta_{ij}.
$$
By $B_{ij}$ we will denote the matrix of $b$-periods of the normalized holomorphic differentials
$$
B_{ij}=\oint_{b_j}\omega_i,\quad i,\, j=1,\dots, g.
$$
Recall that the Jacobi variety (or, simply Jacobian) of $C$ can be realized as a quotient of $\mathbb C^g$ over the lattice of periods of holomorphic differentials
$$
J(C)=\mathbb C^g/\{ 2\pi\sqrt{-1}M+BN\, |\, M,\, N\in \mathbb Z^g\}
$$
Finally, the half-period ${\boldsymbol \varpi}$, for a suitable choice of the basis of cycles (see details in \cite{Fay}) has the form
\beq\label{varpi}
{\boldsymbol \varpi}=\pi\sqrt{-1}(1,0,1,0,\dots)+\frac12\sum_{i=1}^g \left(B_{1i},B_{2i}, \dots, B_{gi}\right).
\eeq
Let
\beq\label{theta}
\theta(\bu)=\sum_{{\bf n}\in \mathbb Z^g} \exp\left\{ \frac12\langle {\bf n}, B{\bf n}\rangle +\langle {\bf n},\bu\rangle \right\}
\eeq
be the Riemann theta-function of the curve $C$ associated with the chosen basis of cycles. Here $\bu=(u_1, \dots, u_g)$ is the vector of independent complex variables, ${\bf n}=(n_1, \dots, n_g)\in\mathbb Z^g$,
$$
\langle {\bf n}, B{\bf n} \rangle=\sum_{j, \, k=1}^g B_{jk}n_j n_k, \quad \langle {\bf n}, {\bf u}\rangle=\sum_{k=1}^g n_k u_k.
$$
Then the tau-function of the solution to the KdV hierarchy corresponding to the the matrix \eqref{2gon} reads
\beq\label{thetakdv}
\tau(\bt)=e^{\frac12\sum_{i,\, j\geq 0}q_{ij}t_i t_j}\theta\left(\sum t_k \bv^{(k)}-\bu_0\right)
\eeq
where $\theta(\bu)$ is the theta-function of the hyperelliptic curve \eqref{hyperell}, the point $\bu_0$ is specified by eqs. \eqref{div1}--\eqref{div3}, the vector $\bv^{(k)}=\left(V^{(k)}_1, \dots, V^{(k)}_g\right)$, $k\geq 0$ is made of the $b$-periods of the normalized second kind differentials
\eqa\label{differs}
&&
\Omega^{(k)} =\frac{2k+1}2 \left( z^{g+k}+c_{k\,1} z^{g+k-1}+\dots+c_{k, g+k}\right) \frac{dz}{w}=
\nn\\
&&
=\frac12\left( (2k+1) z^{k-\frac12} +\sum_{i\geq 0} \frac{q_{k\, i}}{z^{i+\frac32}}\right) \, dz, \quad k\geq 0
\\
&&
\oint_{a_i}\Omega^{(k)}=0, \quad V_i^{(k)}=\oint_{b_i} \Omega^{(k)}, \quad i=1, \dots, g
\nn
\eeqa
and the coefficients $q_{ij}$ come from the regular part of the expansion \eqref{differs}. The vectors $\bv^{(k)}$ can also be computed from the expansions of normalized holomorphic differentials
\beq\label{differs1}
\omega_i=\frac12 \sum_{k\geq 0}V_i^{(k)}\frac{dz}{z^{k+\frac32}},\quad i=1, \dots, g.
\eeq

From Theorem \ref{thm15} we derive

\begin{cor} \label{cor31} The logarithmic derivatives of the theta-function of the hyperelliptic spectral curve \eqref{hyperell} of the matrix polynomial $W(z)$ of the form \eqref{2gon} for any $N\geq 3$ satisfy the following identities
\eqa\label{glav}
&&
\sum_{k_1, \dots, k_N\geq 0} \frac{\pal^N \log \theta\left(\sum t_k \bv^{(k)}-\bu_0\right)}{\pal t_{k_1}\dots \pal t_{k_N}}\Big |_{\bt=0}\frac1{z_1^{k_1+\frac32}\dots z_N^{k_N+\frac32}}=
\nn\\
&&
=-\frac1{N}\frac1{w(z_1)\dots w(z_N)}\sum_{s\in S_N}\frac{\tr\left[  W\left( z_{s_1}\right)\dots W\left( z_{s_N}\right)\right]}{\left(z_{s_1}-z_{s_2}\right)\dots \left( z_{s_{N-1}} -z_{s_N}\right)  \left( z_{s_N}-z_{s_1}\right)}
\eeqa
of formal series in inverse powers of independent variables $z_1^{1/2}$, \dots, $z_N^{1/2}$.
\end{cor}

As an immediate consequence we obtain

\begin{cor} If \eqref{hyperell} is the spectral curve of a matrix polynomial \eqref{2gon} with rational coefficients then the logarithmic derivatives
$$
\frac{\pal^N \log \theta\left(\sum t_k \bv^{(k)}-\bu_0\right)}{\pal t_{k_1}\dots \pal t_{k_N}}\Big |_{\bt=0}
$$
of its theta-function evaluated at the point \eqref{div1}--\eqref{div3} are rational numbers for any $N\geq 3$.
\end{cor}

Generalizations of this Corollary for more general spectral curves will be given in \cite{D18}.


\subsection{Theta-functional approximations of the Witten--Kontsevich tau-function}\par

Let us consider the Airy operator
\beq\label{Ai}
L=\frac{d^2}{dx^2}+2x.
\eeq
Then the matrix resolvent computed in \cite{BDY1} has the form \eqref{wser} with 
\beq\label{Ai1}
a^0_{3k-2}=-\frac12\frac{(6k-5)!!}{24^k(k-1)!}, \quad b^0_{3k}=\frac{(6k-1)!!}{24^k k!}, \quad c^0_{3k-1}=-\frac{6k+1}{6k-1}\frac{(6k-1)!!}{24^k k!},
\eeq
all other coefficients vanish. Recall  \cite{BDY1} that for this solution of the KdV hierarchy the logarithmic derivatives of the tau-function are related to the intersection numbers of the psi-classes in the cohomologies of the Deligne--Mumford moduli spaces of stable algebraic curves
$$
F^0_{i_1\dots i_N}={\prod_{i=1}^N (2k_i+1)!!}\int_{\overline{M}_{g,N} }\psi_1^{k_1}\dots \psi_N^{k_N}
$$
assuming
$$
i_1+\dots+i_N-N+3=3g.
$$

Let us consider for this particular case the procedure of approximation of $\log\tau(\bt)$ by logarithms of theta-functions. As an example consider the 9-truncation of $F(\bt)=\log\tau(\bt)$
\beq\label{wk9}
\left[F(\bt)\right]_9=\frac{t_1}{24}+ 
 \frac{t_4}{1152}+ \frac{t_1^2}{48}+ \frac{t_0t_2}{16}  +\frac16t_0^3 + 
\frac1{72} t_1^3 + \frac1{12}t_0t_1t_2+ \frac1{48}t_0^2 t_3+\frac1{6} t_0^3t_1 +\frac1{6} t_0^3t_1^2 + \frac1{24}t_0^4t_2 .
\eeq
According to the Corollary this polynomial coincides, modulo the linear and quadratic terms with the 9-truncation of the Taylor logarithmic expansion of the genus 4 theta-function
$$
\log\theta\left(t_0 \bv^{(0)}+\frac13 t_1 \bv^{(1)}+\frac1{15} t_2 \bv^{(2)}+\frac1{105} t_3 \bv^{(3)}+\frac1{945}t_4 \bv^{(4)}-\bu_0\right)
$$
of the spectral curve
\beq\label{KW}
w^2+\det W(z)=0, \quad W(z)=\left(\begin{array}{cc} -\frac12 z^3 -\frac{35}{16} & z^4+\frac58 z\\
\\
z^5-\frac78 z^2 & \frac12 z^3+\frac{35}{16}\end{array}\right),
\eeq
see Figure 1. This immediately follows from Corollary \ref{cor31}. To be on the safe side we have checked this statement by a numerical computation of the theta-function. Some hints from \cite{FK} were helpful for us in this computation.

\begin{figure}
\centerline{\includegraphics[width=0.4 \textwidth]{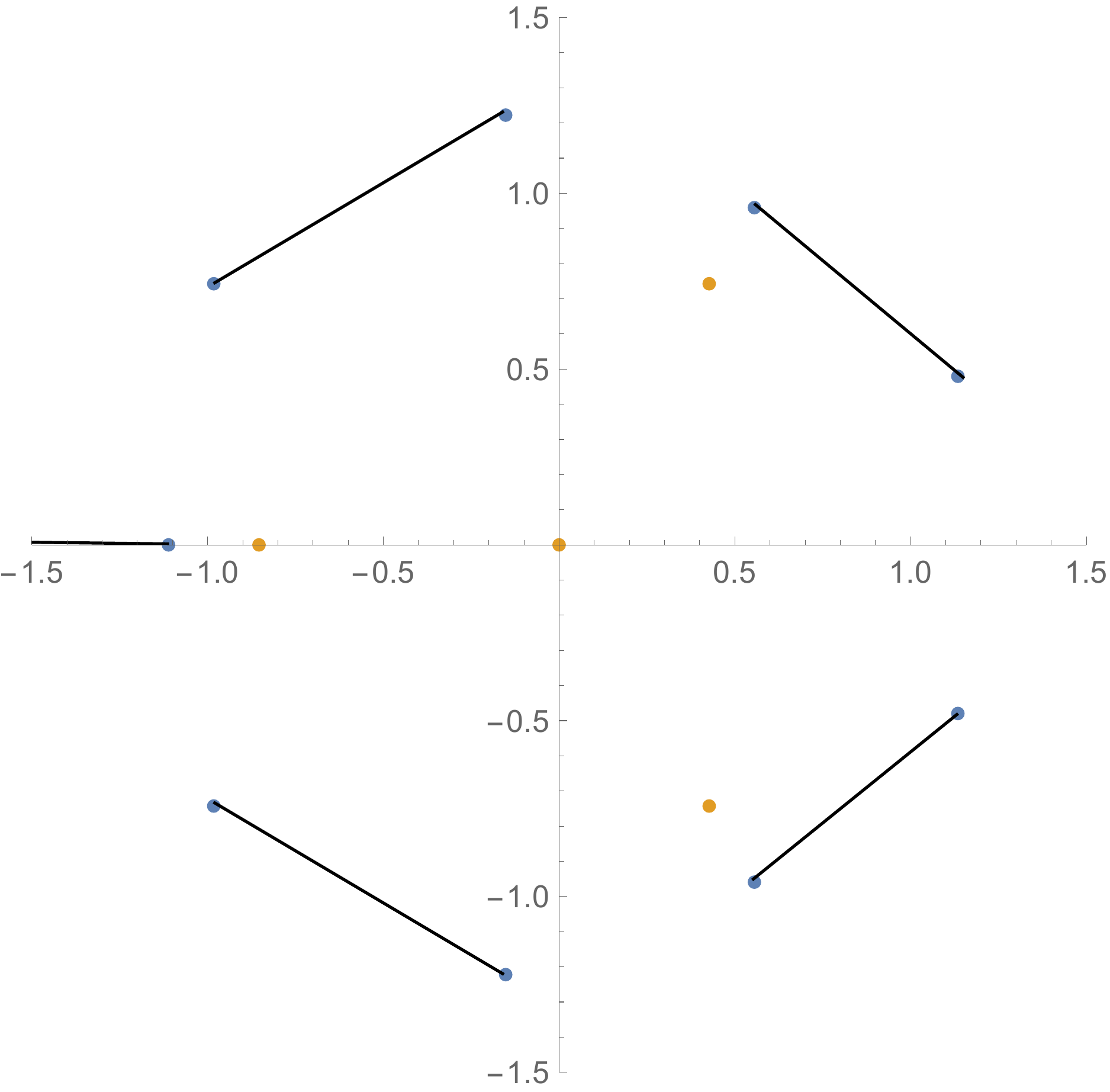}}
\caption{Branchcuts of the spectral curve \eqref{KW}. The yellow dots show the points $\left(0, \frac{35}{16}\right)$, $\left( -\frac{ 5^{1/3}}2, \frac{15}8\right)$, $\left( -e^{2\pi i/3}\frac{ 5^{1/3}}2, \frac{15}8\right)$,  $\left( -e^{-2\pi i/3}\frac{ 5^{1/3}}2, \frac{15}8\right)$ of the divisor $D_0$.} 
\label{zeros}
\end{figure}

Choose the basis of cycles in the following way. The cycles $a_1$, \dots, $a_4$ go around the finite branch cuts. Every cycle $b_i$ crosses once the cycle $a_i$ and also the infinite branch cut. This yields the following matrix of $b$-periods of normalized holomorphic differentials
$$
B=\left(\oint_{b_j}\omega_i\right)=-\left(
\begin{array}{cccc}
 5.800 & 2.811 & 1.720 & 0.895 \\
 2.811 & 7.374 & 3.263 & 1.722 \\
 1.720 & 3.263 & 7.376 & 2.815\\
 0.895 & 1.722 & 2.815 & 5.807 \\
\end{array}
\right)
+\sqrt{-1}\left(
\begin{array}{cccc}
 1.272 & 1.842 & 2.376 & 3.137 \\
 1.842 & 2.116 & 3.137 & 3.898 \\
 2.376 & 3.137& 4.158 & 4.431 \\
 3.137 & 3.898 & 4.431 & 5.000 \\
\end{array}
\right)
$$
where
$$
\omega_1=\left[(-0.866+0.572\, i) z^3+(1.456\, +0.275\, i) z^2-(1.137\, +1.343\, i) z+(0.064\, +1.643 \,
i)
\right]\frac{dz}w
$$
$$
\omega_2=\left[(-1.332+0.261\, i) z^3-(0.260\, -1.042\, i) z^2-(0.316\, -1.270\, i) z+(1.994\, +1.213 \,
i)
\right]\frac{dz}w
$$
$$
\omega_3=\left[(-1.332-0.261\, i) z^3-(0.260\, +1.042\,i) z^2-(0.316\, +1.270\, i) z+(1.994\, -1.213 \,
i)
\right]\frac{dz}w
$$
$$
\omega_4=\left[(-0.866-0.572\, i) z^3+(1.456\, -0.275\, i) z^2-(1.137\, -1.343\, i) z+(0.064\, -1.643 \,
i)
\right]\frac{dz}w.
$$
With the help of \eqref{differs1} we obtain the vectors $\bv^{(k)}$
\eqa
&&
\bv^{(0)}=\left( -1.731 + 1.145 \, i, -2.664 + 0.522 \, i, -2.664 - 
 0.522 \, i, -1.731 - 1.145 \, i\right)
\nn\\
&&
 \bv^{(1)}=\left(2.912 + 0.551 \, i, -0.520 + 2.083 \, i, -0.520 - 
 2.083 \, i, 2.912 - 0.551 \, i\right)
\nn\\
&&
 \bv^{(2)}=\left(-2.273 - 2.685 \, i, -0.632 + 2.541 \, i, -0.632 - 
 2.541 \, i, -2.273 + 2.685\, i\right)
 \nn\\
&&
 \bv^{(3)}=\left(0.127 + 3.286 \, i, 3.987 + 2.426 \, i, 3.987 - 
 2.426 \, i, 0.127 - 3.286 \, i\right)
 \nn
 \eeqa
 and $\bv^{(4)}=0$. Finally, we compute the Abel--Jacobi image
 $$
 \left(4.506 + 5.841 \, i, 6.826 + 1.741 \, i, 6.826 - 
 1.741 \, i, 4.506 - 5.841 \, i\right)
 $$
  of the divisor $D_0$ and the point $\bu_0\in J(C)\setminus (\Theta)$
  $$
  \bu_0=\left( 10.119 - 1.614 \, i, 14.411 - 3.756 \, i, 14.413 - 
 11.933 \, i, 10.126 - 14.074 \, i\right).
 $$
 To avoid computational problems with big real parts it is convenient to shift this point by a vector of the lattice
 $$
 \bu_0\mapsto \bu_0+B(1,1,1,1) .
 $$
 Finally, we arrive at the following 9-truncated expansion
 \eqa
 &&
 \left[ \log\theta\left(t_0 \bv^{(0)}+\frac13 t_1 \bv^{(1)}+\frac1{15} t_2 \bv^{(2)}+\frac1{105} t_3 \bv^{(3)}+\frac1{945}t_4 \bv^{(4)}-\bu_0\right)\right]_9=
 \nn\\
 &&
 =0.447+0.002\,i -(0.577+0.003\,i)t_0+0.186\,t_1+0.05\, t_2
 \nn\\
 &&
+ 0.228\, t_0^2+0.175\, t_0 t_1+0.07\, t_1^2+0.091 \, t_0 t_2-0.057\, t_1 t_2-0.016\, t_0 t_3
\nn\\
&&
+0.167\, t_0^3+0.014\, t_1^3+0.084 \, t_0 t_1 t_2+0.021\, t_0^2 t_3+0.167\, t_0^3 t_1+0.167\, t_0^3 t_1^2+0.042 \, t_0^4 t_2.
 \nn
 \eeqa
We have omitted terms less than $10^{-3}$. Two small imaginary numbers in the first line come from some numerical errors to be settled. The first two lines are of no interest but the third line satisfactory matches the expansion \eqref{wk9} (the apparent discrepancy in the coefficient in front of $t_0 t_1 t_2$ is $\simeq 5\cdot 10^{-4}$).

\begin{figure}
\centerline{\includegraphics[width=0.4 \textwidth]{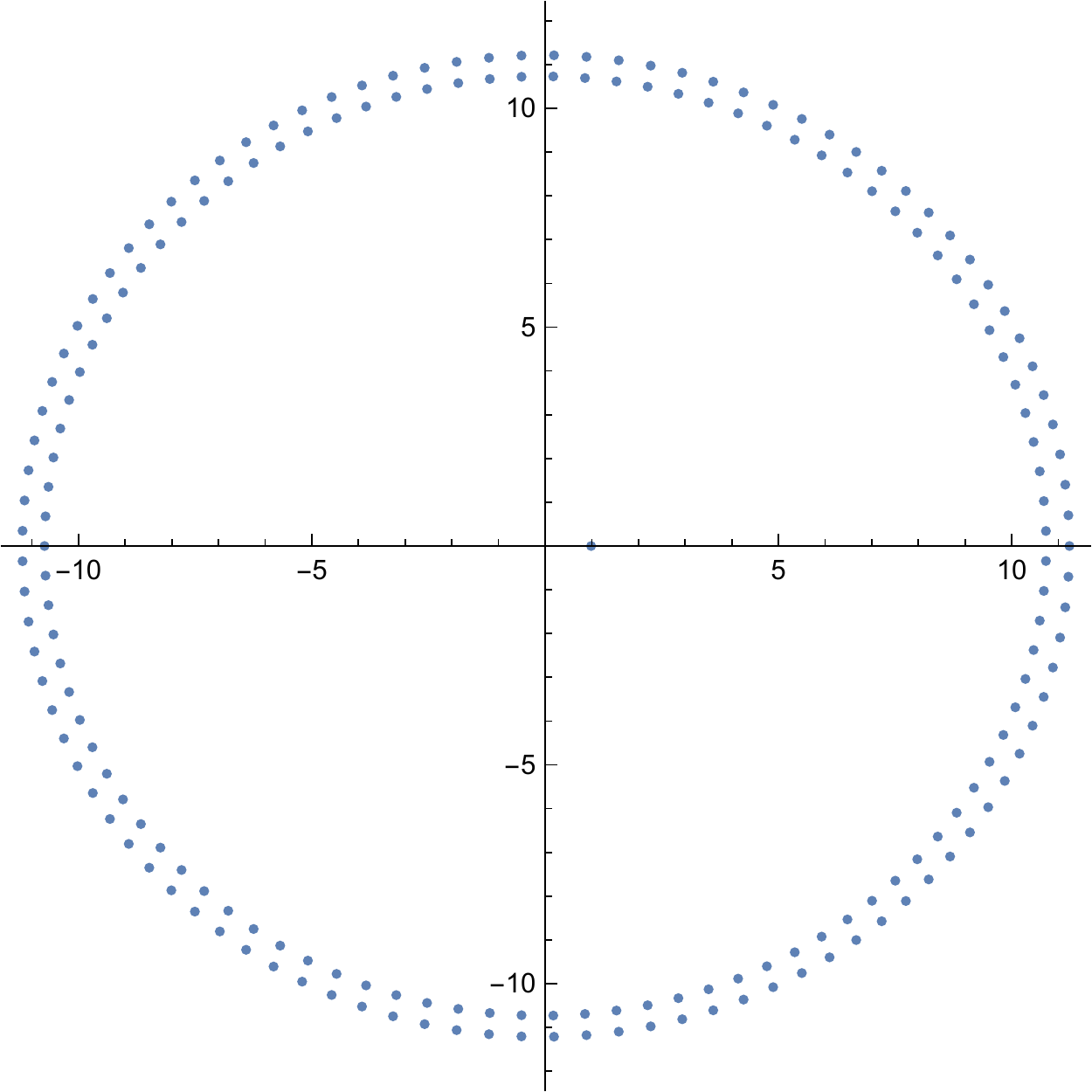}}
\caption{Branch points of the spectral curve of the matrix $W^0_{100}(z)$} 
\label{zeros100}
\end{figure}

Let us also make few comments about the theta-functional approximations of the Witten--Kontsevich tau-function for growing order of the truncation. For a given $g>0$ denote
$$
W^0_g(z)=\left(\begin{array}{cc} 0& 1\\z+c_1^0 & 0\end{array}\right)+\sum_{i= 1}^g\left(\begin{array}{cc} a_i^0 & b_i^0 \\ c_{i+1}^0  & -a_i^0 \end{array}\right)\frac1{z^i}.
$$
In order to compute the $m$-truncated $\log\tau(\bt)$ with $m=2g$ or $m=2g+1$ according to Corollary one has to deal with the spectral curve of $W^0_g(z)$. Here we will consider one particular subfamily of spectral curves with $g=3n+1$. In that case there are $6n+3$ zeroes of the equation $\det W^0_{3n+1}(z)=0$ that, for large $n$ are distributed in the following way. One of them is on the real axis near $z=1$. Other zeros are located near two circles (see Figure \ref{zeros100} for $n=33$). On the inner circle there are $3n$ zeros and on the outer one there are $3n+2$ zeroes. The inner zeros are close to roots of the equation\footnote{Recall that the roots of this equation are the $z$-projections of the poles of the eigenvector of $W^0_{3n+1}(z)$.}
$$
1+\sum_{i=1}^{3n}\frac{b_i^0}{z^i}=0
$$
while the outer poles are close to the roots of
$$
z+\sum_{i=1}^{3n+2}\frac{c_i^0}{z^{i-1}}=0.
$$
Taking into account only the leading terms of these equations one obtains the following asymptotics for the radii of these two circles for large $n$
$$
\log R_{\rm in}\sim \frac23 (\log 3n -1)-\frac{\log n}{6n}+{\mathcal O}\left(\frac1n\right), \quad \log R_{\rm out}\sim \frac23 (\log 3n -1)+\frac{\log n}{18n}+{\mathcal O}\left(\frac1n\right)
$$
so
$$
\frac{R_{\rm out}}{R_{\rm in}}\sim n^{\frac2{9n}}.
$$

The structure of the spectral curve for $g=3n+2$ is identical to the above one as $W^0_{3n+2}(z)=W^0_{3n+1}(z)$. For $g=3n$ the spectral curve has a double point at $z=0$. The branch points are still located near two circles, like for the above case $g=3n+1$ but there is no branch point near $z=1$. Appearance of this exceptional real branch point for $g=3n+1$ requires a better understanding.

{\small \it SISSA, Via Bonomea, 265, Trieste, Italy}

\end{document}